# Acceleration of the Boundary Element Method for arbitrary shapes with the Fast Fourier Transformation

JUSTUS BENAD

Technische Universität Berlin, Berlin 10623, Germany
E-Mail: mail@jbenad.com
Orcid ID: 0000-0002-8677-3260

This work illustrates the possibility to apply the Fast Fourier Transformation to obtain the integrals of the Boundary Element Method (BEM) on arbitrary shapes. The procedure is inspired by the technique used with great success within the framework of the half-space approximation in contact mechanics. There, the boundary integral equations are given by simple convolutions over the boundary surface. For arbitrary shapes this is not the case. Thus, the FFT and the great reduction in computational complexity that comes with it cannot be utilized as easily. In this work, it is illustrated that although the integral equations of the BEM are not convolutions over boundary of arbitrary shapes, they are indeed convolutions over the space which is one dimension higher than that of the boundary. Therefore, the FFT can indeed be used to calculate the BEM integral equations on arbitrary shapes, only this comes at the cost of increasing the dimension of the FFT. A small example is given which illustrates how the concept can be used to fully solve a BEM problem with a given closed boundary using only the FFT approach to obtain the integral equations and no hybrid techniques or other approximations.

## 1. Introduction

In the past decades, the Boundary Element Method (BEM) has been used for the analyses of various engineering problems [1]. For some applications, it has become a strong alternative to the Finite Element Method (FEM). For example, the boundary element method has become a standard tool in contact mechanics [2, 3] where it is used with great success within the framework of the half-space approximation (see for example [4, 5]). For a half-space, the integral equations of the boundary element method are simply given by a two-dimensional convolution over the surface which can easily be obtained with the Fast Fourier Transformation (FFT). For a half-space surface with $n \times n$ discretization points on a uniform grid, the computational complexity is $O(n^2 \log n)$ when the FFT is used. When it is not used and the entire $n^2 \times n^2$ matrix of the linear system of equations is built, the complexity is $O(n^6)$ using direct solvers or $O(n^4)$ using iterative solvers.

For arbitrary shapes, the integral equations of the boundary element method are not given by simple convolutions over the boundary surface as it is the case for the half-space. Thus, the FFT and the reduction in computational complexity that comes with it cannot be used as easily for this case as it is possible for the case of the half-space. For a cube with $n$ discretization points on each side ($6n^2$ surface points), the matrix of the linear system of the BEM has $6n^2 \times 6n^2$ entries. As above, when no measures are taken to accelerate the calculation, the complexity is $O(n^6)$ with direct solvers or $O(n^4)$ using iterative solvers. Many valuable techniques have been proposed over the past years to reduce this complexity (see for example [6, 7, 8, 9])

This work seeks to illustrate and highlight the following: Although the integral equations of the BEM are not convolutions over boundary of arbitrary shapes, they are indeed convolutions over the space which is one dimension higher than that of the boundary. Therefore, the FFT can indeed be used to calculate the BEM integral equations on arbitrary shapes, only then the dimension of the FFT has to be increased.

It will be shown with an example that this realization can be used to fully solve a BEM problem with an arbitrary boundary using only the FFT to obtain the integral equations.

## 2. Boundary integral formulations of the governing equations

Let us use LAPLACE's and NAVIER's equation to illustrate how the boundary integrals of the BEM can be calculated on arbitrary shapes with the FFT. LAPLACE's equation is

$$\Delta u(\underline{x}) = 0 . \qquad (1)$$

Adding a point source in $\underline{x}_0$ yields the equation

$$\Delta u^*(\underline{x}) = -\delta(\underline{x} - \underline{x}_0) . \qquad (2)$$

The solution of (2) is the *fundamental solution* of (1). For example, it is

$$u^*(\underline{x} - \underline{x}_0) = -\frac{1}{2\pi} \ln(|\underline{x} - \underline{x}_0|) \qquad (3)$$

for the two-dimensional case of $\underline{x} = x\underline{e}_x + y\underline{e}_y$. With the weak formulation of (1), the definition of the fundamental solution (2), and the divergence theorem, one can obtain the boundary integral formulation of (1) as

$$c(\underline{x}_0)u(\underline{x}_0) = \int_S q(\underline{x})u^*(\underline{x} - \underline{x}_0)dS \\ - \int_S u(\underline{x})q^*(\underline{x} - \underline{x}_0)dS \qquad (4)$$

which relates the values for $u(\underline{x})$ and $q(\underline{x}) = \underline{n}(\underline{x}) \cdot \nabla u(\underline{x})$ on the boundary $S$ of a region ($\underline{n}$ is the outward normal vector) to the value of $u$ at a particular point $\underline{x}_0$ of the region. It is $c = 1$ when $\underline{x}_0$ lies inside the region, and $c = 1/2$ when $\underline{x}_0$ lies on the smooth boundary $S$. Note that for the two-dimensional case with $\underline{x} = x\underline{e}_x + y\underline{e}_y$ the boundary $S$ is a line, whereas for the three-dimensional case with $\underline{x} = x\underline{e}_x + y\underline{e}_y + z\underline{e}_z$ the boundary $S$ is a surface.

The boundary integral formulation of NAVIER's equation

$$\frac{1}{1-2\nu}\nabla(\nabla \cdot \underline{u}(\underline{x})) + \Delta \underline{u}(\underline{x}) = -\frac{1}{\mu}\underline{b}(\underline{x}) \quad (5)$$

shall be introduced in a similar way. Here we consider only the three-dimensional case with $\underline{x} = x\underline{e}_x + y\underline{e}_y + z\underline{e}_z$. Choosing the volume force field $\underline{b}$ with a delta function to represent a point force acting in $\underline{x}_0$ in the direction of a unit vector $\underline{m}$ yields

$$\frac{1}{1-2\nu}\nabla(\nabla \cdot \underline{u}(\underline{x})) + \Delta \underline{u}(\underline{x}) = -\frac{1}{\mu}\delta(\underline{x}-\underline{x}_0)\underline{m}. \quad (6)$$

The solution to (6) is called fundamental solution of (5). It is given through

$$\underline{\underline{u}}^*(\underline{x}-\underline{x}_0) = \frac{\frac{3-4\nu}{|\underline{x}-\underline{x}_0|}\underline{\underline{I}} + \frac{(\underline{x}-\underline{x}_0)\otimes(\underline{x}-\underline{x}_0)}{|\underline{x}-\underline{x}_0|^3}}{16\pi\mu(1-\nu)} \quad (7)$$

via $\underline{u}^* = \underline{\underline{u}}^* \cdot \underline{m}$. With $\underline{u}^*$, one can now obtain the boundary integral formulation of (5) as

$$c(\underline{x}_0)\underline{u}(\underline{x}_0) = \int_S \underline{t}(\underline{x}) \cdot \underline{u}^*(\underline{x}-\underline{x}_0)dS \\ - \int_S \underline{u}(\underline{x}) \cdot \underline{t}^*(\underline{x}-\underline{x}_0)dS, \quad (8)$$

where $c$ is defined as above, $\underline{u}$ and $\underline{t} = \underline{\underline{\sigma}} \cdot \underline{n}$ are the values on the boundary $S$ and $\underline{t}^*$ is known from $\underline{u}^*$ with the material law.

## 3. Application of the FFT

Now let us take a closer look at the two boundary integral formulations from above, (4) and (8). Both strongly remind of the integral formulation one obtains for an elastic half-space [3] and which can be solved rapidly with the FFT as it represents a two-dimensional convolution. It reads

$$u_a(x,y) = \int_S K_{ab}(x-x_0, y-y_0)\sigma_b dx_0 dy_0, \quad (9)$$

where $u$ is the deformation, $\sigma$ is the load, $a$ is the direction of the deformation and $b$ is the direction of the load, $(a,b) \in \{x,y,z\}$. Note that in (9), the two coordinates $x$ and $y$ lie in the plane of the surface $S$ over which the integration is performed. Thus, the uniform grid of the FFT which is used to calculate the convolution in (9) is perfectly aligned with the half space surface. (see Fig. 1).

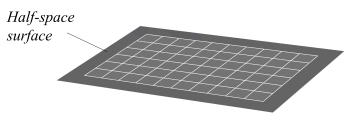

**Fig. 1** A uniform two-dimensional grid aligned with the surface of an elastic half-space

Although this seems trivial, it is precisely this characteristic which is missing in (4) and (8). In both of these boundary integral formulations, the structure is that of a cross-correlation which is equal to the convolution due to the symmetry of the fundamental solutions. However, the integration is not, as in (9), performed over a boundary which corresponds to the entire parameter space, but instead over a boundary which lies *within* this parameter space. For example, in the two-dimensional case of the LAPLACE equation, the integration has to be performed over a line. If the integration would have to be performed over the entire surface enclosing this line, it would be a convolution. The same holds true for the three-dimensional NAVIER equation. Here the integration over the boundary is an integration over a surface lying arbitrarily within the three-dimensional space. If the integration would have to be performed over the entire volume enclosing this surface, it would be a convolution.

This observation leads to the realization, that if one simply does perform the convolution over the entire parameter space enclosing the boundary, and not only over the boundary itself, one can indeed calculate the integrals of the BEM such as in (4) and (8) with the FFT. One only has to pay close attention to appropriately setting zeros wherever there is no boundary such as not to distort the results which are of interest (see the next section for an example). One can almost think of this procedure as if one obtains the results on a thin boundary shell and then divides them by the shell thickness to obtain the final results on the boundary. Fig. 2 displays an arbitrary shape through which a uniform grid is placed on which the FFT can be performed.

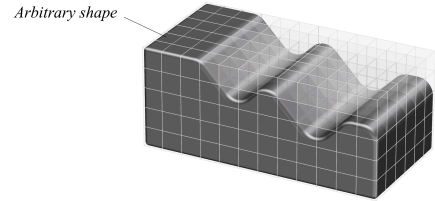

**Fig. 2** An arbitrary three-dimensional shape fully enclosed with a uniform three-dimensional grid

Naturally, this procedure can only be performed at a certain cost, which is the additional dimension of the FFT. Let us consider a two-dimensional example for which the boundary integral equations have to be calculated on a line. Consider a square with $n$ discretization points on each side. The matrix of the linear system of the BEM has $4n \times 4n$ entries. Without any measures to accelerate the calculation, the computational complexity to invert this matrix is $O(n^3)$ with direct solvers or $O(n^2)$ using iterative solvers. When the FFT is used to obtain the integrals as proposed above, the entire two-dimensional space in which the line lies has to be discretized. The computational complexity to perform the two-dimensional FFT is then $O(n^2 \log n)$. Thus, there is no reduction in computational complexity for the two-dimensional case. Let us now consider the three-dimensional example of a cube with $n$ discretization points on each side ($6n^2$ surface points). Here, the matrix of the linear system of the BEM has $6n^2 \times 6n^2$ entries. When no measures are taken to accelerate the calculation, the computational complexity is $O(n^6)$ with direct solvers or $O(n^4)$ using iterative solvers. With the FFT though, the complexity is only $O(n^3 \log n^{1.5})$. This reduction in complexity makes the use of the FFT appealing. Moreover, highly efficient implementations for the FFT exist and the algorithm can be run on parallel systems such as the Graphics Processing Unit (GPU).

## 4. Example for the Laplace equation

We now present a small example which should be considered as a *very rough sketch* of how to use the method described above to solve a BEM problem with an arbitrary boundary with the FFT. Consider again the two-dimensional form of LAPLACE's equation (1). As boundary, we choose a unit square which is discretized with $n$ points on each side. A certain potential $u$ shall be fixed on the boundary as it is shown with the blue line in Fig. 3. The goal of the operation is to obtain $q$ on the boundary, characterizing the outward flow.

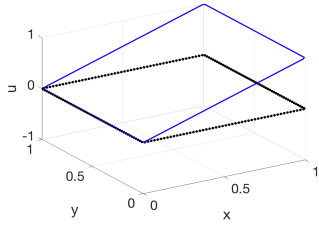

**Fig. 3** A unit square with a fixed potential (blue line) on the boundary. On each side of the square there are $n$ discretization points (black dots).

Let us insert the fundamental solution (3) of LAPLACE's equation into the boundary element formulation (4). Introducing

$$r = \sqrt{(x-x_0)^2 + (y-y_0)^2}, \quad (10)$$

the fundamental solution (3) can be written as

$$u^*(x-x_0, y-y_0) = -\frac{1}{2\pi}\ln(r). \quad (11)$$

With (11) and $q = \underline{n} \cdot \nabla u$, we then obtain

$$q^* = -\frac{\underline{n} \cdot \left((x-x_0)\underline{e}_x + (y-y_0)\underline{e}_y\right)}{2\pi r^2}. \quad (12)$$

Both (11) and (12) can now be inserted into (4) which turns to

$$c(x_0,y_0)u(x_0,y_0)$$
$$= -\frac{1}{2\pi}\int_S \ln(r) q \, dS \quad (13)$$
$$+ \frac{1}{2\pi}\int_S u \frac{(x-x_0)n_x + (y-y_0)n_y}{r^2} dS.$$

Note that for this example the integration in (13) is along a line which lies in the two-dimensional space. Note also that the values for the components of the outward normal vector $\underline{n}$ depend on the vector's position on the border: $n_x = n_x(x,y)$, $n_y = n_y(x,y)$. Let us now split up the second integral in (13) and write out the dependencies in more detail:

$$c(x_0,y_0)u(x_0,y_0)$$
$$= -\frac{1}{2\pi}\int_S \ln\left(r(x-x_0, y-y_0)\right) q(x,y) dS$$
$$+ \frac{1}{2\pi}\int_S \frac{(x-x_0)}{r(x-x_0, y-y_0)^2} u(x,y) n_x(x,y) dS \quad (14)$$
$$+ \frac{1}{2\pi}\int_S \frac{(y-y_0)}{r(x-x_0, y-y_0)^2} u(x,y) n_y(x,y) dS.$$

One can now see, that each one of the three integrals in (14) would be a *two dimensional cross correlation*, if it was taken over a surface. However, all three integrals are taken only over the line. In a similar way, the three integrals in (14) - if one imagines them again to be taken over the surface and not the line - could also be interpreted as a *two dimensional convolution*. One only has to rewrite the equations in the following way:

$$c(x_0,y_0)u(x_0,y_0)$$
$$= -\frac{1}{2\pi}\int_S \ln\left(r(x_0-x, y_0-y)\right) q(x,y) dS$$
$$- \frac{1}{2\pi}\int_S \frac{(x_0-x)}{r(x_0-x, y_0-y)^2} u(x,y) n_x(x,y) dS \quad (15)$$
$$- \frac{1}{2\pi}\int_S \frac{(y_0-y)}{r(x_0-x, y_0-y)^2} u(x,y) n_y(x,y) dS.$$

In the example, $u$ is given on the boundary. To apply the convolution, $u$ is now extended to the entire parameter space by adding zero elements wherever there is no boundary (see Fig. 4). Also, the parameter space is extended in the appropriate fashion to apply the FFT.

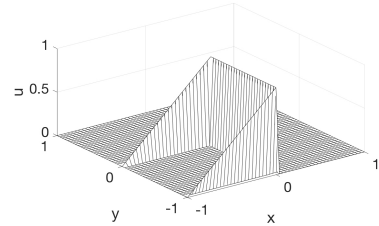

**Fig. 4** Zero elements are added surrounding the given values for $u$ on the boundary and the parameter space is extended in the appropriate fashion to apply the FFT

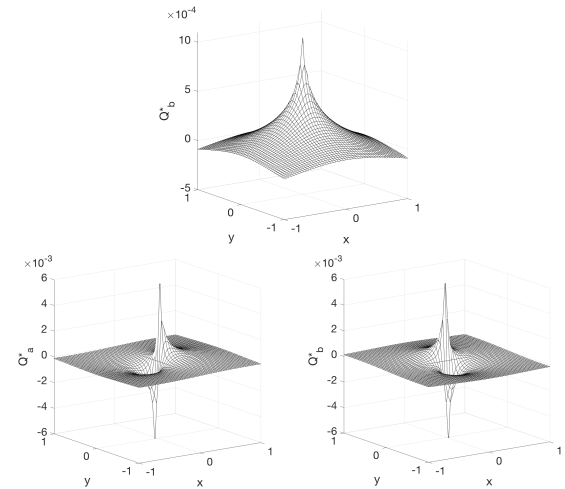

**Fig. 5** Rough approximation of the three kernel functions which are required for the example

Let us now assume that the integrals in (15) are evaluated over the entire parameter space. The three kernel functions of the first, second and third integral, which shall be referred to as $U^*$, $Q_a^*$ and $Q_b^*$, are roughly approximated on constant elements on the entire parameter space which is extended for the application of the FFT (see Fig. 5). The second and third extended integral can then be obtained. For example, one simply has

$$\text{IFFT}(\text{FFT}(u_{\text{extended},x}) * \text{FFT}(Q_a^*)) \qquad (16)$$

for the second extended integral. Therein $u_{\text{extended},x}$ are the values of $u$ extended with zeros on the entire parameter space as shown in Fig. 4. The $x$ in the index denotes that the values on the boundary are further multiplied with the $x$-portion of the normal vector $\underline{n}$. The results of (16) are of course values over the entire parameter space. The values which are of interest lie only on the boundary line and have to be divided by the discretization length to obtain the actual results of the integral.

The values of $q$ which we seek cannot be calculated directly as they lie within the first integral in (15). We therefore use a conjugant gradient method. Therein, the first integral is evaluated in the same style as the other two with the use of the FFT. After an initial guess $q$ is changed in each iteration until the norm of the residual is below a predefined value. The result for $q$ is displayed with the red line in Fig. 6.

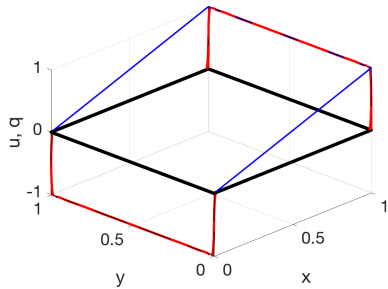

**Fig. 6** Visualization of the results for the outward flow $q$ (red) on the boundary for the given distribution of the potential $u$ (blue) on the boundary. The results were generated using the FFT to obtain the integral equations as it is described above. For the displayed results it was $n = 500$, and 5 iterations were required with the conjugant gradient method.

## 5. Summary

In this work it was illustrated that the boundary integral formulations of the LAPLACE and NAVIER equation can be regarded as convolutions over the space which is one dimension higher than that of the boundary of an arbitrary shape. Therefore, the FFT can be used to calculate the BEM integral equations on arbitrary shapes in a similar manner like it has become a standard within the framework of the half-space approximation generally used in contact mechanics. The flat shape of the half-space is perfectly aligned with a two-dimensional grid over which a convolution can be performed. For arbitrary shapes, a grid has to be introduced which fully runs through and encloses the surface. This makes the use of the FFT possible but comes at the cost of having to add a dimension to the FFT. However, there is a reduction in computational complexity when compared to the conventional technique of inverting the fully populated BEM matrix. For a cube with $n$ discretization points on each side, the matrix of the linear system of the BEM has $6n^2 \times 6n^2$ entries. When no measures are taken to accelerate the calculation, the computational complexity is $O(n^6)$ with direct solvers or $O(n^4)$ using iterative solvers. With a three-dimensional FFT though, the complexity is only $O(n^3 \log n^{1.5})$. This reduction in complexity makes the use of the FFT appealing. Moreover, highly efficient implementations for various use cases exist for the FFT.

At the end of the paper it was briefly illustrated with a small example how the concept can be used to fully solve a BEM problem with a closed boundary using only the FFT approach to obtain the integral equations and no hybrid techniques or other approximations.

The author suspects that the concept may be a key to developing efficient boundary element solvers in the near future.

**Acknowledgement:** The author would like to thank V. L. Popov for many valuable discussions on the topic and critical comments.